# Balancing Variety and Sample Size: Optimal Parameter Sampling for Ariel Target Selection

Emilie Panek[1], Alexander Roman[1], Katia Matcheva[1], Konstantin T. Matchev[1], and Nicolas B. Cowan[2,3]
[1] Department of Physics and Astronomy, The University of Alabama, Tuscaloosa, AL 35487, USA; epanek1@ua.edu
[2] Department of Earth and Planetary Sciences, McGill University, 3450 University St., Montréal, H3A 2A7, Canada
[3] Department of Physics, McGill University, 3600 University St., Montréal, QC H3A 2T8, Canada
*Received 2026 January 28; revised 2026 July 16; accepted 2026 July 17; published 2026 August 7*

## Abstract

Targeted astrophysical surveys are limited by the amount of telescope time available, which makes it difficult to observe every single object of interest. In order to maximize the scientific return, we need a well thought-out strategy for selecting the observational targets, in our case exoplanets. This study evaluates various strategies for selecting exoplanet targets within limited observation windows, focusing specifically on the selection of exoplanet targets for Tier 2 transit spectroscopy with ESA's upcoming Ariel mission. We define three distinct selection criteria for the selected sample—sample size, variance, and leverage—and translate them into objective functions compatible with modern optimization algorithms. The performance of these methods is demonstrated through three practical exercises across one, two, and three parameters of diversity. Each of the three objective functions represents a unique trade-off between sample size, diversity, and total observation time. While a time-greedy approach maximizes the quantity of planets, it fails to capture diversity. Conversely, variance-greedy selection prioritizes diversity but introduces significant drawbacks: it oversamples rare cases and undersamples typical planets, ultimately reducing the total number of targets observed. Leverage-based selections emerge as the most effective middle ground, successfully balancing sample diversity with a robust sample size. Specifically, when maximizing sample leverage, we test five heuristics: "leverage greedy," "simulated annealing," "*K*-means clustering," "regular classes," and "quantile classes." This work supports the broader community effort to ensure that Ariel delivers the most diverse and scientifically valuable sample of exoplanet atmospheres within mission limits.

## 1. Introduction

The Ariel mission is a space telescope led by ESA, planned to launch in 2029 (G. Tinetti et al. 2018, 2021). It will operate in synergy with ongoing and planned missions such as JWST (J. P. Gardner et al. 2006, 2023), Twinkle (B. Edwards et al. 2019; I. Stotesbury et al. 2022; T. Zhang et al. 2025), Pandora (T. Barclay et al. 2025), and PLATO (H. Rauer et al. 2025). Ariel is uniquely designed for population-level studies of exoplanet atmospheres rather than single-object characterization. Its primary science goal is to characterize the atmospheres of a large and diverse population of exoplanets in order to understand their formation, evolution, and chemical composition. It aims to obtain spectral information for close to 1000 transiting exoplanets, which would allow a significant understanding of the physics and chemistry of a wide range of systems to be built. Observations are organized in a tiered strategy: an initial large sample is observed for basic atmospheric characterization (Tier 1), followed by a smaller subset of targets observed with a higher signal-to-noise ratio (SNR) to enable more detailed constraints on atmospheric composition (Tier 2). With more than 6000 confirmed transiting exoplanets,[4] one is forced to make certain selection choices to maximize the science return of the telescope within its 4 yr primary mission. The problem of optimal target selection for Ariel and similar missions has received increasing attention in recent literature. It has been realized that selecting a subsample optimized for individual atmospheric studies does not guarantee the best constraints on the global population parameters (N. E. Batalha et al. 2023). N. B. Cowan & B. Coull-Neveu (2025) pointed out that the precision with which we can quantify linear population trends is determined by the number of targets $N$ in the survey and their variance $V$. The authors discussed the notion of leverage, namely $L = \sqrt{NV}$, as a compromise between the multiplicity and diversity of the observed planets. We note that this definition differs from the classical notion of leverage in regression analysis: here, it refers to a quantity defined over the entire selected sample and is used as an objective function. The choice of $L$ was motivated by the fact that the uncertainty on the inferred slope of a linear trend is inversely proportional to the leverage. Subsequent work in J. A. Burt et al. (2025) emphasized the importance of sampling more evenly across a range of planet radii and equilibrium temperatures. It also considered the additional costs in terms of ground-based telescope time to reduce the uncertainties in the orbital planet parameters.

The specific problem considered here—optimal target selection subject to a finite observational budget—falls into a general class of problems in optimization theory known as submodular maximization with a knapsack constraint. Submodular maximization implies a property of diminishing returns, where adding an element to a small subset helps more than adding it to a large subset. A knapsack constraint means that each selected element $i$ comes with a cost (in this case this is the required observational time $t_i$ by Ariel) and that there exists an overall budget which should not be exceeded (in our case, the total lifetime $T$ of the Ariel mission). Thus we can write the constraint as

$$\sum_{i=1}^{N} t_i \leqslant T, \quad (1)$$

where $N = |\mathcal{S}|$ is the number of elements in the selected subsample $\mathcal{S}$ of observational targets.

[4] See https://exoplanetarchive.ipac.caltech.edu/ for the up-to-date number of confirmed exoplanets (J. L. Christiansen et al. 2025).

A key step is the choice of an objective function $f(\mathcal{S})$ that reflects the scientific merit of the selected subsample $\mathcal{S} \subseteq \mathcal{T}$, where $\mathcal{T}$ spans the entire available set of planets. To be specific, in this paper we focus on the Ariel Mission Candidate Sample (MCS; B. Edwards & G. Tinetti 2022) list, that defines the entire set $\mathcal{T}$ which we can choose from. Following N. B. Cowan & B. Coull-Neveu (2025), in this paper we shall primarily adopt the leverage as our objective function, but for illustration we shall also consider alternative choices for $f$ (sample size and sample variance), since different choices of $f$ would result in different optimal selection outcomes for $\mathcal{S}$.

After settling on an objective function $f$, it is still necessary to pick the set of planetary parameters $\{p\}$ that will be used as axes of diversity. Common choices include the planetary radius $R_p$, the planet equilibrium temperature $T_p$, and sometimes the stellar effective temperature $T_s$ (B. Edwards & G. Tinetti 2022; N. E. Batalha et al. 2023; B. J. Hord et al. 2024; J. A. Burt et al. 2025; N. B. Cowan & B. Coull-Neveu 2025). In our study we shall consider those three parameters as well, i.e., $\{p\} = \{R_p, T_p, T_s\}$. We note that our method can be applied to any other set of planetary parameters of interest, but we decided to use these three for consistency with previous studies (B. Edwards & G. Tinetti 2022; B. J. Hord et al. 2024; N. B. Cowan & B. Coull-Neveu 2025).

The resulting numerical task of selecting a subset $\mathcal{S}$ which maximizes the objective function $f(\{p\})$ subject to the budgetary constraint (1) is in general far from trivial. A brute force approach (considering all possible candidate selections $\mathcal{S}$) is impossible in practice since, in the case of Ariel, this is an astronomical number of combinations, on the order of $10^{650}$ (N. B. Cowan & B. Coull-Neveu 2025). This is why one tends to rely on simple heuristics or figures-of-merit for each planet in order to arrive at an approximate solution. For example, N. B. Cowan & B. Coull-Neveu (2025) developed a class-based approach by binning along each parameter axis, thus dividing the set $\mathcal{T}$ into different categories (thus aiming for diversity) and then cyclically choosing the easiest-to-observe planet in each category. While sensible, this heuristic does not guarantee that the selected subsample $\mathcal{S}$ achieves the global maximum of $f$. In this paper we propose a leverage-greedy algorithm which directly targets the optimization of the objective function $f = L$ and is much more likely to yield a globally optimal selection for $\mathcal{S}$. We also consider several alternative selection strategies for the case $f = L$. We compare our leverage-greedy method to the previous approaches in the literature and demonstrate that it gives a superior selection in terms of leverage. Moving beyond the specific cases examined here (such as $f = L$), this work addresses the general methodological challenge of target selection by introducing a framework applicable to a wide range of science objectives.

The paper is organized as follows. Section 2 introduces the Ariel MCS dataset $\mathcal{T}$ that we will be working with. Section 3 defines three different objective functions and discusses corresponding heuristics used in their implementation. Section 4 contains our numerical results and compares the different outcomes in selecting the mission target list $\mathcal{S}$. Section 5 discusses the advantages and limitations of our approach.

## 2. Data

The MCS for the Ariel mission was defined in B. Edwards & G. Tinetti (2022). It is a list of potential Ariel targets derived from currently known exoplanets and TESS planetary candidates. It is regularly updated and is publicly available on the Ariel mission GitHub page.[5] The catalog is primarily built from the NASA Exoplanet Archive (J. L. Christiansen et al. 2025) using the "Planetary Systems Composite Data" table, which combines measurements from multiple studies to provide a more complete set of planetary and stellar parameters. Missing quantities are inferred where possible following standard relations in the literature (M. J. Pecaut & E. E. Mamajek 2013; J. Chen & D. Kipping 2017), and only planets with sufficient information for Ariel SNR predictions are retained (see L. V. Mugnai et al. 2020 for more information on ArielRad simulations). For this study we used the update from 2025 August 18, which is the latest update available at the time of writing. It includes both confirmed planets and TESS planetary candidates, resulting in 2696 objects. For brevity, we will refer to the objects in that sample as "planets," regardless of whether they have been confirmed yet. We are following the B. Edwards & G. Tinetti (2022) philosophy of lumping together transits and eclipses and we focus on planets that necessitate no more than 20 visits to build up Tier 2 SNR. For Tier 2 the requirement is SNR $\geqslant$ 7 on spectroscopic features in the spectrum (B. Edwards & G. Tinetti 2022), which reduces the initial MCS list to 1343 planets. We also removed an odd candidate with a planet radius larger than 10 Jupiter radii, obtaining a final sample of 1342 planets. This sample is illustrated in Figure 1 as scatter plots of the required Tier 2 observational time versus each of the planet parameters in $\{p\} = \{R_p, T_p, T_s\}$. The histograms on each axis are projections onto the corresponding marginal distributions.

The observation time of each planet transit is taken as three times the actual transit duration. While the usual guideline is to adopt 2.5 times the transit duration to account for the baseline needed before and after the transit, we use a factor of 3 to include an additional margin for telescope slew and settling time. For the purposes of this paper, we consider a three year Ariel Tier 2 transit survey ($T = 3$ yr).

## 3. Methods

In this section we introduce three examples for the objective function $f(\mathcal{S})$ and in each case discuss the computational strategy for selecting the optimal subsample $\mathcal{S}$. For the leverage objective function $f_L(\mathcal{S})$ in Section 3.3, we consider five different computational approaches (heuristics) for the sample selection.

### 3.1. Sample Size (Time-greedy Selection)

One simple choice is to maximize the number of observed planets:

$$f_t(\mathcal{S}) = |\mathcal{S}| = N, \quad (2)$$

[5] https://github.com/arielmission-space/Mission_Candidate_Sample

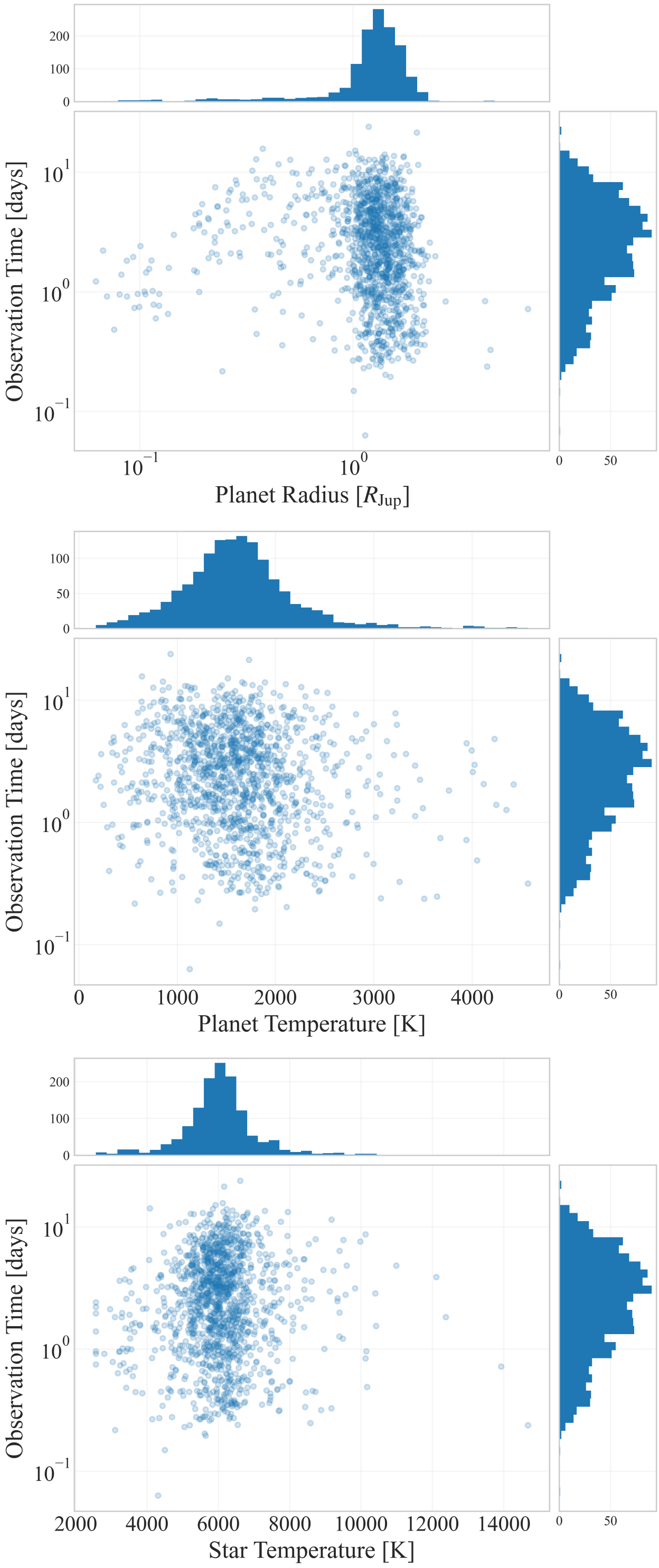

**Figure 1.** Scatter plots of the total observation time per planet required to reach an Ariel Tier 2 transit SNR versus the planet radius (top panel), the planet temperature (middle panel), and the star temperature (bottom panel). The histograms are projections onto the corresponding marginal distributions.

where the vertical bars denote the cardinality of the set $\mathcal{S}$, i.e., the number of elements $N$ in $\mathcal{S}$. This is a very conservative choice, which makes sense in cases where it is not immediately obvious (or there is no consensus agreement) what are the important parameters $\{p\}$ to diversify. The overall time budget constraint (1) implies that this choice of $f$ prioritizes the planets that need the least observation time, hence the index $t$ in (2). In this scenario, the numerical procedure of selecting the optimal $\mathcal{S}$ is very simple: we rank-order all planets in our dataset by their observation time $t_i$ and apply a time-greedy selection, starting from the easiest-to-observe planet, and continuing until the total available observation time $T$ is exhausted. The resulting selection is equivalent to the single class method described in N. B. Cowan & B. Coull-Neveu (2025).

### 3.2. Sample Variety (Variance-greedy Selection)

At the other extreme, one could prioritize diversity of the sample along some chosen parameter $p$ and thus define the objective function to be the corresponding variance:

$$f_v(\mathcal{S}) = \frac{1}{N}\sum_{i\in\mathcal{S}}(p_i - \bar{p}_{\mathcal{S}})^2 \equiv \frac{1}{N}\sum_{i\in\mathcal{S}} v_i, \tag{3}$$

where

$$\bar{p}_{\mathcal{S}} \equiv \frac{1}{N}\sum_{i\in\mathcal{S}} p_i \tag{4}$$

is the average value of the parameter $p$ within the selected subsample $\mathcal{S}$ and $v_i$ denotes the contribution of planet $i$ to the sum in (3), hence the subscript $v$. As before, $N = |\mathcal{S}|$ denotes the number of elements in $\mathcal{S}$.

To select the subsample which maximizes (3), we can follow a similar greedy selection procedure as in the previous subsection, ordering the planets by their respective $v_i$. However, there is an important subtlety: the mean $\bar{p}_{\mathcal{S}}$ is not known a priori and so we cannot compute the squared deviations $v_i = (p_i - \bar{p}_{\mathcal{S}})^2$ exactly. At the same time, for a large enough subsample $\mathcal{S}$, we can reasonably assume that the sample mean is very close to the population mean

$$\bar{p}_{\mathcal{T}} \equiv \frac{1}{|\mathcal{T}|}\sum_{i\in\mathcal{T}} p_i, \tag{5}$$

which is known to us in advance. We can thus make the approximation $\bar{p}_{\mathcal{S}} \approx \bar{p}_{\mathcal{T}}$ and rank-order the planets by the squared deviations $v_i = (p_i - \bar{p}_{\mathcal{T}})^2$ instead, starting the selection from the largest $v_i$ and proceeding until the total available observation time $T$ is exhausted. In what follows, we shall refer to the resulting selection as "variance greedy."

### 3.3. Sample Leverage

A major science goal of Ariel is to quantify population-level trends in atmospheric properties. The simplest trends are linear: $y = mp + b$, where $y$ represents a derived planetary property measured with Ariel, while $p$ is our previously discussed planetary parameter along an axis of diversity (D. P. Thorngren et al. 2016; L. Welbanks et al. 2019; M. R. Swain et al. 2024). The constants $m$ and $b$ need to be extracted from a fit to the data collected from the sample $\mathcal{S}$ of observational targets. Therefore, the choice of $\mathcal{S}$ impacts the subsequent precision with which $m$ and $b$ could be determined. N. B. Cowan & B. Coull-Neveu (2025) pointed out that the uncertainty $\sigma_m$ on the fitted slope $m$ is inversely proportional to the leverage $L$ and, therefore, $L$ is a sensible choice for the objective function $f$. In order to avoid dealing with square roots, in this work we find it more convenient to introduce the

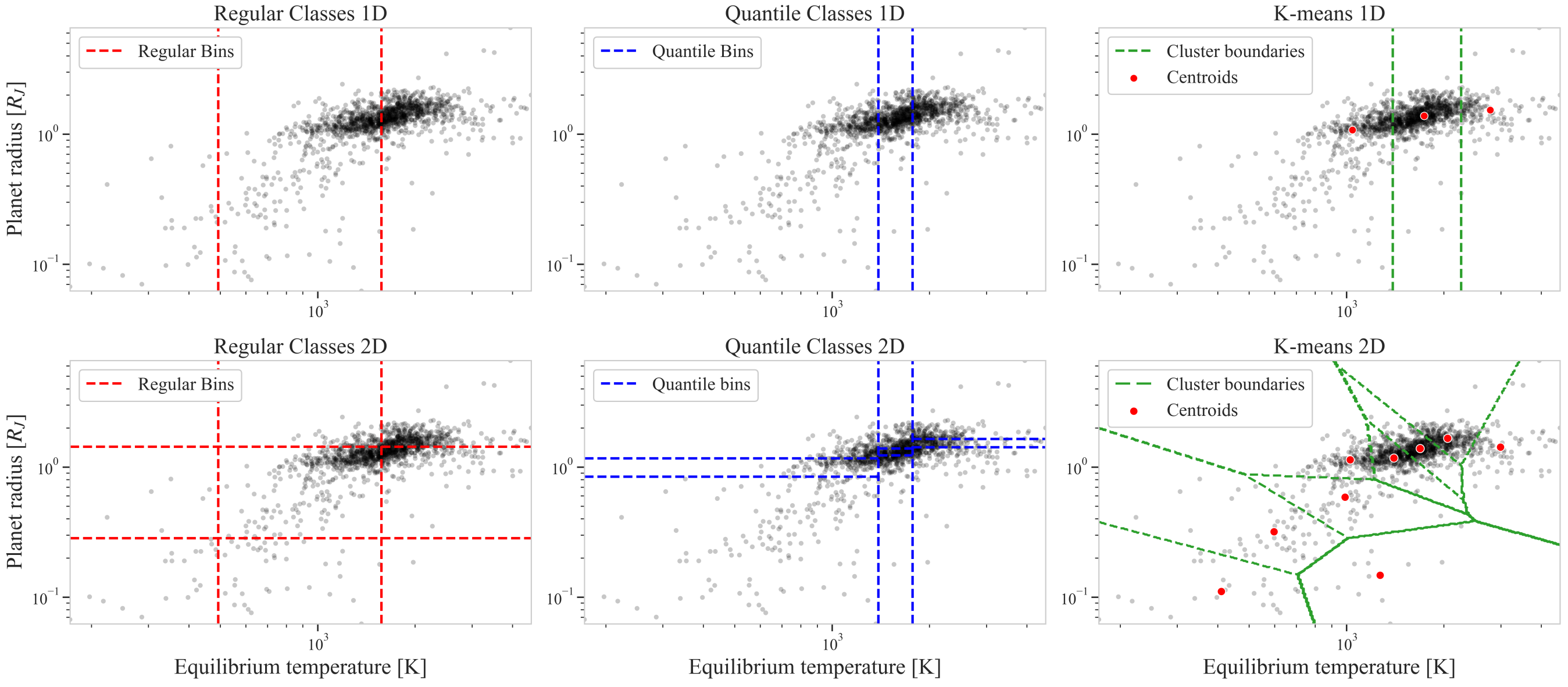

**Figure 2.** Partitioning of the parameter space for the regular-classes, quantile-classes, and $K$-means clustering methods. The bins are shown for the 1D (top row) and 2D (bottom row) cases. The 3D case with 27 bins is too cluttered to be added in this figure. The number of planets in each bin resulting from these three methods is given in Figure 3.

leverage squared:

$$f_L(\mathcal{S}) = L^2(\mathcal{S}) = N \cdot \mathrm{Var}(p) = \sum_{i \in \mathcal{S}} (p_i - \bar{p}_{\mathcal{S}})^2 = \sum_{i \in \mathcal{S}} v_i, \qquad (6)$$

where $\bar{p}_{\mathcal{S}}$ and $v_i$ are the same quantities appearing in Equation (3). The objective function $f_L$ strikes a balance between sample size (through the $N$ factor) and diversity (through the Var($p$) factor).

The mathematical problem of choosing the optimal subsample $\mathcal{S}$ which maximizes the objective function (6) is rather nontrivial. N. B. Cowan & B. Coull-Neveu (2025) proposed a couple of heuristic methods (Sections 3.3.1–3.3.2) which for reference we include in our numerical analysis in the next section. In this paper, we consider three additional methods (Sections 3.3.3–3.3.5), among which the "leverage-greedy" approach of Section 3.3.4 is designed to provide the best possible answer. For completeness, we shall also illustrate what happens if the selection is done totally at random, ignoring any objective function $f$ (Section 3.4).

### 3.3.1. Regular Classes

To broaden the variety of studied exoplanets, they are typically divided into specific groups, such as hot Jupiters, warm sub-Neptunes, lava planets, etc., and then the target selection is done by picking the easiest-to-observe target from each group, alternating between the classes until the observation budget is full (B. Edwards & G. Tinetti 2022; B. J. Hord et al. 2024). This strategy relies on an appropriate definition of the planetary classes. Following N. B. Cowan & B. Coull-Neveu (2025), here we define the classes directly by binning the parameter space into either regular or quantile bins, without referencing the planet type.

With regular binning, the relevant range along each axis of diversity is divided into a fixed number of equal-length intervals. In practice, this is done by dividing the range of each parameter into a fixed number of intervals and assigning each planet to a class based on its value along that axis. The number of planets per class is therefore determined by the underlying distribution of the sample. N. B. Cowan & B. Coull-Neveu (2025) considered up to 20 intervals and showed that splitting into so many is typically unnecessary, since the highest leverage was observed in selections done with just a handful of classes. For concreteness, in our numerical experiments in the next section, we shall adopt the default of three classes used in N. B. Cowan & B. Coull-Neveu (2025). The bins defined by this method in one dimension (top) and two dimensions (bottom) are represented in the left panels of Figure 2 and the corresponding number of planets in each bin is shown in the left panels of Figure 3.

### 3.3.2. Quantile Classes

Since planet populations tend to be not uniformly spread along the axes of diversity, regular binning runs into the problem of insufficient counts in certain sparsely populated bins. As a result, the selection algorithm may reach a point where there are no planets left in a class to choose from. To avoid this problem, N. B. Cowan & B. Coull-Neveu (2025) also considered using quantile bins along the axes of diversity, so that each class contains the same number of planets. In this case, the bin boundaries are chosen such that each class contains approximately the same number of objects, ensuring that all classes remain populated throughout the selection process (compare the panels in the left and middle columns in Figures 2 and 3). Unlike N. B. Cowan & B. Coull-Neveu (2025), in the 2D and 3D cases the bins for the next axis of diversity are defined within each bin along the previous axis of diversity, ensuring uniform coverage in the multidimensional cases. As before, the algorithm selects the cheapest (in terms of required observation time) planet, alternating between the classes until the total budget $T$ is exhausted. Following N. B. Cowan & B. Coull-Neveu (2025), we shall illustrate this scenario also with three quantile bins per axis.

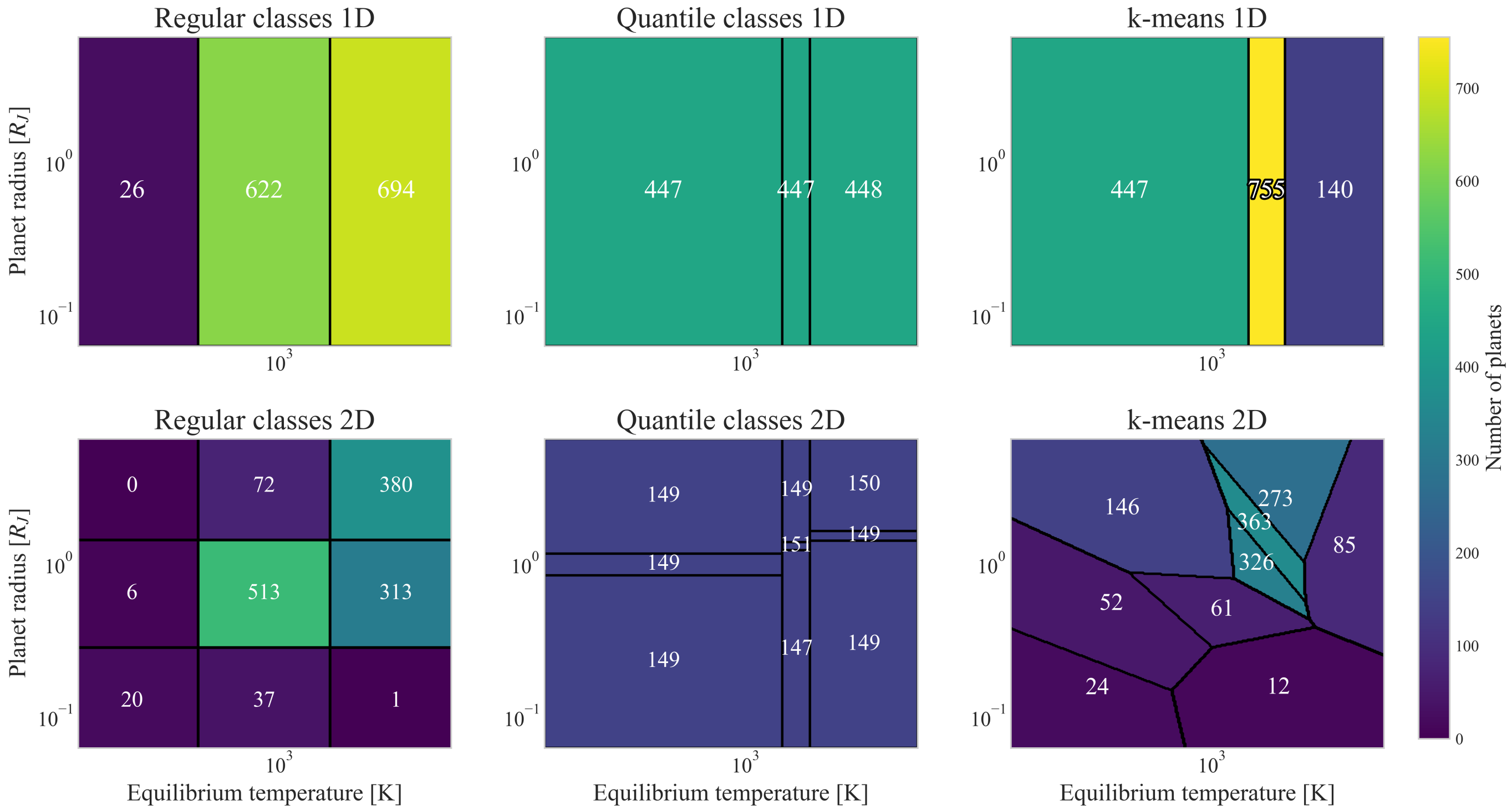

**Figure 3.** The number of planets per bin resulting from the three binning methods shown in Figure 2.

### 3.3.3. K-*means Clustering*

An alternative method of separating planets into classes is to use unsupervised machine learning clustering techniques. For example, $K$-means clustering finds a predefined number of $K$ clusters of planets in the data (J. J. C. Hayes et al. 2020; K. T. Matchev et al. 2022). Each planet is assigned to the nearest cluster center in parameter space, which results in partitioning the sample into $K$ groups. Note, however, that the number of planets that end up in a given cluster is unconstrained and is determined by the clustering algorithm. The selection algorithm then again cycles between the clusters, selecting the cheapest remaining one in each cluster, until the budget is exhausted. This approach automatically adapts to a potentially uneven distribution of planets, since the planets within each cluster have similar properties, while planets belonging to different clusters generally have different properties. The number of planets in each cluster is not fixed a priori and depends on the structure of the dataset. For a fair comparison with the class-based methods from Sections 3.3.1 and 3.3.2, we shall use three clusters per axis (for a total of 9 in 2D and 27 in 3D). The bins defined by this method are shown in the right column of Figure 2 and the corresponding number of planets in each bin is given in the right panels of Figure 3.

### 3.3.4. *Leverage-greedy Selection*

Here we introduce our fiducial greedy algorithm. For submodular maximization of the objective function $f_L$ with a knapsack constraint (1) this algorithm works as follows. Let the current selection be $\mathcal{S}_0$. For each of the remaining planets $e_i \in \mathcal{T}$ whose selection would not cause a violation of the time budget, i.e., for which the following inequality holds

$$t_i + \sum_{k \in \mathcal{S}_0} t_k \leqslant T, \tag{7}$$

compute the marginal gain per unit cost

$$\frac{\Delta f_L(e_i)}{\Delta t} = \frac{f_L(\mathcal{S}_0 \cup e_i) - f_L(\mathcal{S}_0)}{t(\mathcal{S}_0 \cup e_i) - t(\mathcal{S}_0)} = \frac{f_L(\mathcal{S}_0 \cup e_i) - f_L(\mathcal{S}_0)}{t_i} \tag{8}$$

and proceed iteratively with a greedy selection. As the name suggests, the ratio (8) quantifies the benefit (in terms of the objective function $f_L$) from adding a single element $e_i$ with a cost $t_i$ to the existing selection $\mathcal{S}_0$, keeping in mind the budgetary constraint. Using the definition (6), we can express the marginal gain simply as

$$\frac{\Delta f_L}{\Delta t} = \frac{v_i}{t_i} = \frac{(p_i - \bar{p}_{\mathcal{S}_0})^2}{t_i}, \tag{9}$$

where in analogy to (4), $\bar{p}_{\mathcal{S}_0}$ is the average value of the parameter $p$ over the set $\mathcal{S}_0$. Since the values of $p_i$ and $t_i$ for each planet are known in advance, and the average $\bar{p}_{\mathcal{S}_0}$ can be updated at the start of each iteration, Equation (9) provides a simple ranking criterion for our "leverage-greedy" method, which balances both variance and number of planets as it prioritizes planets with the highest leverage score (9).

Note that in cases where the size of the optimal subsample $\mathcal{S}$ is large enough to justify the approximation $\bar{p}_{\mathcal{S}_0} \approx \bar{p}_{\mathcal{T}}$ at all times, we can use an alternative criterion in terms of the population average $\bar{p}_{\mathcal{T}}$,

$$\frac{\Delta f_L}{\Delta t} = \frac{v_i}{t_i} = \frac{(p_i - \bar{p}_{\mathcal{T}})^2}{t_i}, \tag{10}$$

which leads to a further simplification since it allows the marginal gains (10) to be precomputed once and for all.

In conclusion of this subsection, let us motivate the marginal gain ratio as the right variable to use with the following simple gedanken experiment. As a starting point, consider a randomly selected subsample $\mathcal{S}_0$ which saturates the time budget constraint in the sense that

$$\sum_{k\in\mathcal{S}_0} t_k \simeq T, \tag{11}$$

so that adding any other additional planet would violate the inequality (7). Since this $\mathcal{S}_0$ was chosen at random, it is far from optimal, and we can improve the leverage by adding a judiciously chosen planet $i$ from the remaining set in $\mathcal{T}$. However, in order to stay under budget, we have to simultaneously eliminate one of the existing planets, say $j$, which is already in $\mathcal{S}_0$. As a result of this swap between planets $i$ and $j$, the objective function (6) changes by

$$\Delta f_L = v_i - v_j. \tag{12}$$

However, now we need to take into account the relative observational costs $t_i$ and $t_j$. If the removed planet has a larger cost, this effectively frees up enough observing time to accommodate more than one planet with cost $t_i$. In other words, removing a planet with cost $t_j$ allows, in principle, the inclusion of up to $t_j/t_i$ planets of type $i$ (assuming an infinite reservoir $\mathcal{T}$). This will keep the budget the same and lead to a change in the objective function

$$\Delta f_L = \frac{t_j}{t_i} v_i - v_j = t_j\left(\frac{v_i}{t_i} - \frac{v_j}{t_j}\right), \tag{13}$$

which makes it clear that the benefit depends only on the ratio $v/t$, in agreement with Equations (8)–(10). This shows that an optimal swap corresponds to replacing elements with low $v/t$ by elements with higher $v/t$, which justifies the use of the marginal gain per unit cost, Equation (10), as the ranking criterion.

#### *3.3.5. Simulated Annealing*

Simulated annealing is a flexible search method designed to explore the space of possible subsamples beyond the purely greedy choices described in (9) or (10). Starting from an initial feasible subset, we generate trial moves by randomly adding, removing, or swapping planets, with the resulting subset remaining within the total observing time budget. Each proposed move is evaluated based on its effect on the leverage objective function $f_L$. In order to avoid getting stuck in local maxima, the annealing algorithm is allowed in principle to accept worse solutions within some tolerance. The algorithm keeps track of a variable (the temperature $\Theta$) that controls how likely it is for worse changes to be accepted. If the proposed change improves the leverage, it is always accepted. If it does not improve the leverage, it is accepted with probability

$$P = \exp\left(\frac{L(\text{new}) - L(\text{current})}{\Theta}\right),$$

where $L$ is the leverage and $\Theta$ is the temperature. Over the course of the optimization the temperature decreases, making the algorithm more conservative. Unlike the leverage-greedy algorithm, simulated annealing does not rely on marginal gains per unit cost and therefore provides an independent consistency check on the quality of greedy solutions, at the expense of increased computational cost.

### *3.4. Random Selection*

To understand how the heuristic methods above compare to random selection, we also sample random subsets uniformly from all possible subsets that (almost) saturate the time constraint. They provide a useful baseline for Figures 4–6 and represent the result from blind selection, without applying any optimization.

### *3.5. Multiple Axes of Diversity*

While our previous discussion was focused on a single parameter of interest, it can be easily generalized to multiple parameters of diversity. To be specific, let us represent the set of parameters of interest $\{p^{(1)}, p^{(2)},\ldots, p^{(N_p)}\}$ with the vector $\boldsymbol{p}$. In analogy to (5), let us define the population mean

$$\langle \boldsymbol{p}\rangle_{\mathcal{T}} \equiv \frac{1}{|\mathcal{T}|}\sum_{i\in\mathcal{T}} \boldsymbol{p}_i\,. \tag{14}$$

Then all of our previous discussion still holds, by replacing $p$ with $\boldsymbol{p}$ and $\bar{p}$ with $\langle \boldsymbol{p}\rangle_{\mathcal{T}}$, e.g.,

$$f_L(\mathcal{S}) = N\cdot \text{Var}(\boldsymbol{p}) = \sum_{i\in\mathcal{S}}(\boldsymbol{p}_i - \langle \boldsymbol{p}\rangle_{\mathcal{S}})^2 \simeq \sum_{i\in\mathcal{S}}(\boldsymbol{p}_i - \langle \boldsymbol{p}\rangle_{\mathcal{T}})^2. \tag{15}$$

Note that when we write out (15) explicitly,

$$f_L(\mathcal{S}) = \sum_{i\in\mathcal{S}}(\boldsymbol{p}_i - \langle \boldsymbol{p}\rangle_{\mathcal{T}})^2 = \sum_{k=1}^{N_p}\sum_{i\in\mathcal{S}}(p_i^{(k)} - \langle p^{(k)}\rangle_{\mathcal{T}})^2, \tag{16}$$

we see that the sum involves terms with different physical units, which may therefore have very different magnitudes. In such cases, the optimization algorithm will focus on the parameters with the largest numerical values (in their respective physical units) and will largely ignore the others. To prevent this, it is common in machine learning to "standardize" each parameter $p^{(k)}$ by centering around its population mean $\langle p^{(k)}\rangle_{\mathcal{T}}$ and rescaling by its standard deviation $\sigma_{\mathcal{T}}^{(k)}$:

$$p_i^{(k)} \rightarrow \hat{p}_i^{(k)} \equiv \frac{p_i^{(k)} - \langle p^{(k)}\rangle_{\mathcal{T}}}{\sigma_{\mathcal{T}}^{(k)}}. \tag{17}$$

In what follows, in multiparameter cases we shall use these standardized variables $\hat{p}^{(k)}$.

## 4. Results

The numerical results from our experiments are summarized in Table 1, while Figures 4–10 illustrate the planetary target selections resulting from the different methods discussed in Section 3. We perform three different exercises: a one-parameter exercise, where the diversity parameter is chosen to be the planet equilibrium temperature (top section of Table 1); a two-parameter exercise, where the parameters of interest are the planet radius and equilibrium temperature (middle section of Table 1); and a three-parameter exercise, where the parameters of interest are the planet radius, the planet

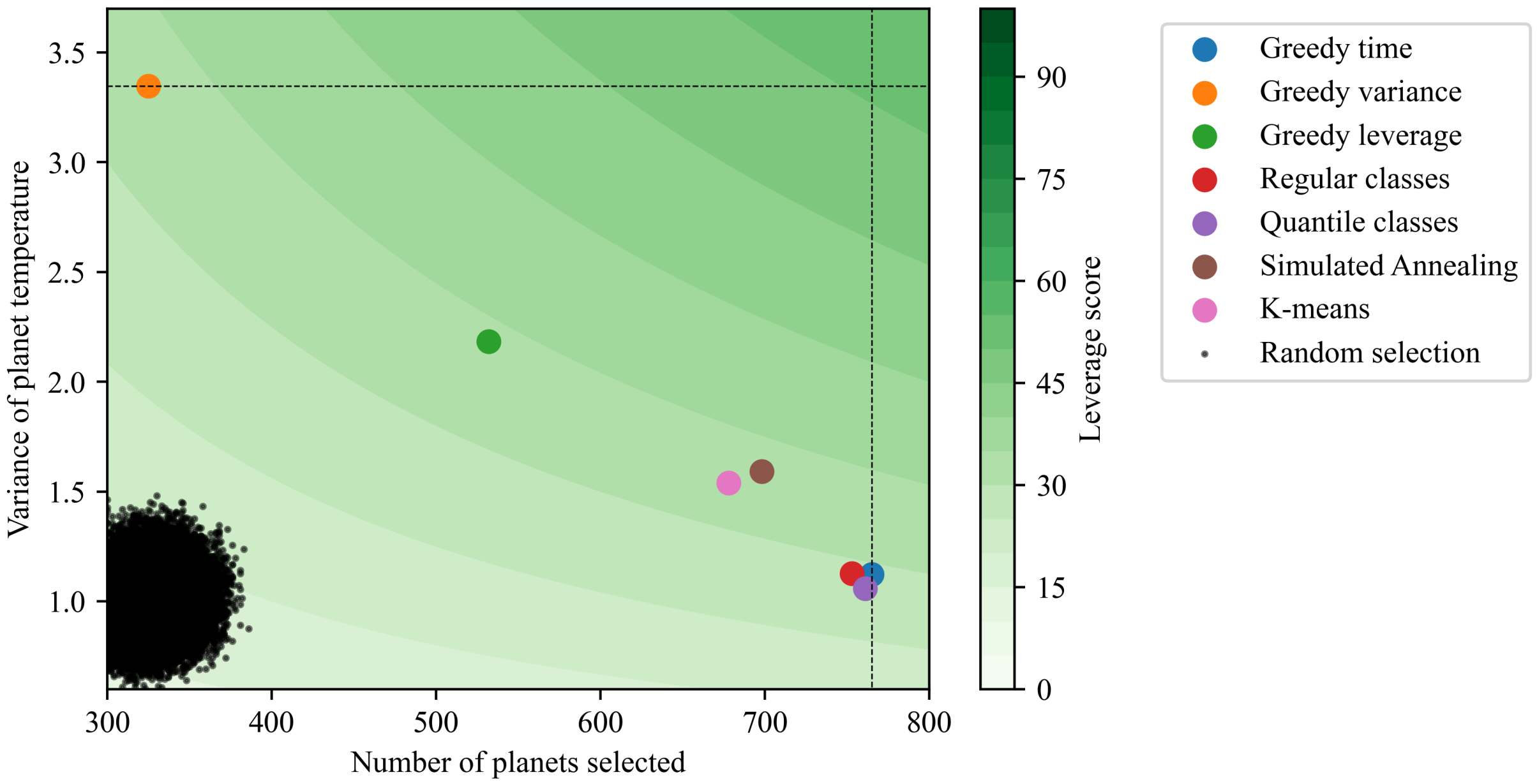

**Figure 4.** One-parameter exercise. A summary plot of the performance of all the different selection methods discussed in the paper for the one-parameter exercise. Colored circles correspond to the seven selection strategies shown in Table 1, while black dots represent randomly chosen subsets. The background is shaded according to the leverage of the selected subset.

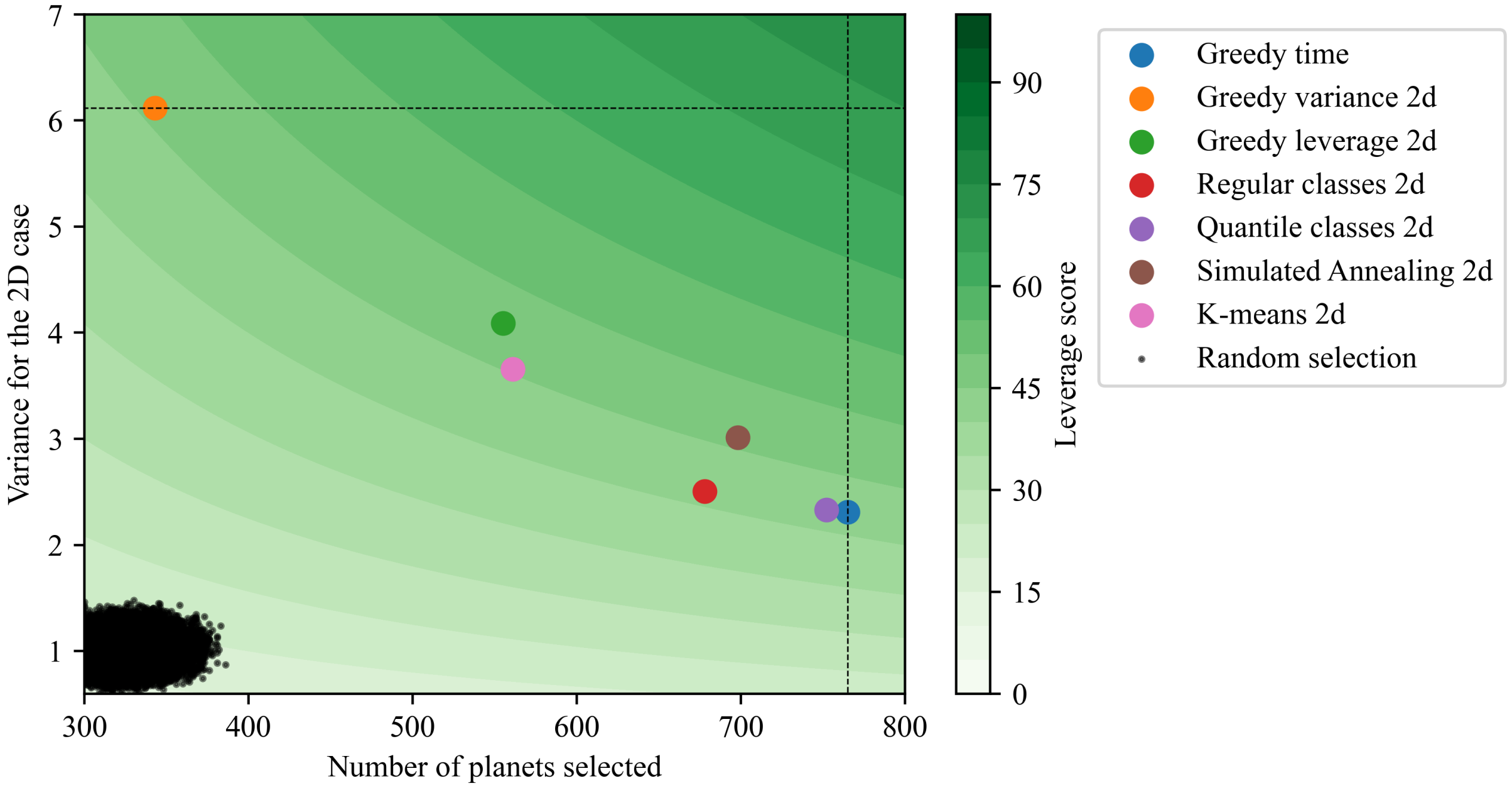

**Figure 5.** Two-parameter exercise. The same as Figure 4, but for the two-parameter exercise.

equilibrium temperature, and the effective temperature of the star (bottom section of Table 1).

In each exercise, we consider the three different objective functions defined in Equations (2), (3), and (6), and report the respective results in Table 1. The five different heuristics for maximizing the leverage objective function (6) are shown in italics. Not all of these heuristics explicitly optimize the leverage objective function, some (regular classes, quantile classes, and $K$-means clustering) instead promote diversity through structured sampling and are included for comparison. For completeness, the results for random selections of planets subject only to the time constraint are also included in the table and are seen to be significantly inferior to the remaining methods. The numerical differences in Table 1 can be understood in terms of how each method samples the underlying parameter distributions. Methods such as variance greedy preferentially select targets in the tails of the distributions, maximizing variance but undersampling the central population. The time-greedy and class-based methods concentrate more strongly around the core of the distributions, producing larger samples but lower leverage. We observe that in each exercise, the leverage-greedy algorithm (third line in each section of Table 1) outperforms all other methods in terms of maximizing the leverage. This is expected; methods that do not explicitly optimize leverage generally underperform compared to the leverage-greedy approach.

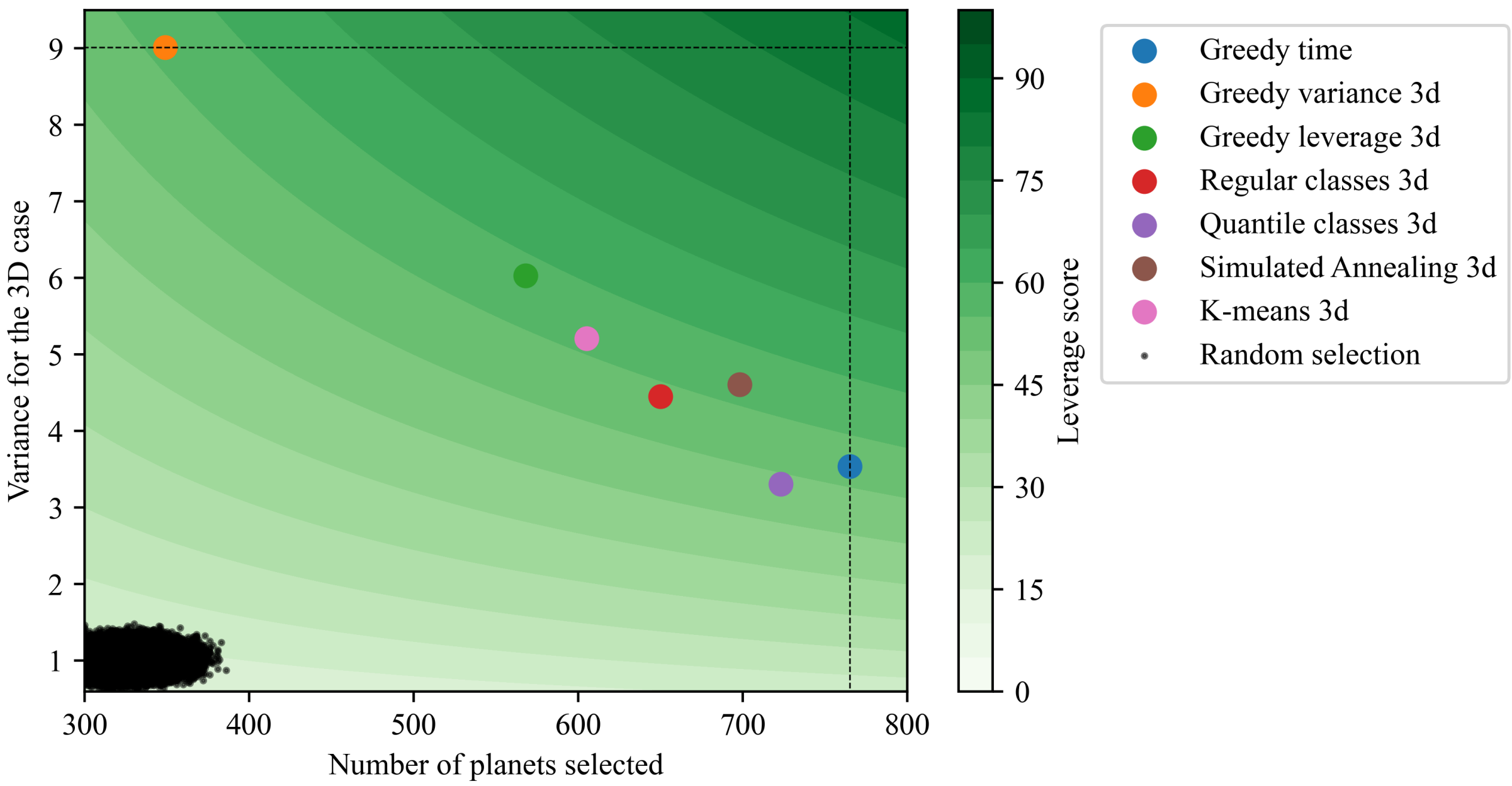

**Figure 6.** Three-parameter exercise. The same as Figure 4, but for the three-parameter exercise.

Each panel in Figures 7 and 8 shows the entire sample of 1342 planets (blue points) as a scatter plot in the plane of observation time versus planet equilibrium temperature. The selected planets are overplotted in orange. The green shading in Figure 7 indicates the contribution of an individual planet to the total time budget $T$ (top panel) or to the respective objective function (middle and lower panels). This color coding helps understand how greedy algorithms work—they start the selection from the most beneficial region and gradually progress through the adjacent green bands until the time budget is exhausted. For example, in the top panel of Figure 4, the selection proceeds from bottom to top, starting with the planets that are easiest to observe. Similarly, in the middle panel of Figure 7, the selection starts from the outliers on the left and right sides (having the largest variance $v_i = (T_i - \langle T \rangle)^2$) and progresses toward the middle. In the bottom panel of Figure 7, the selection starts from the planets with the largest variance per unit time, $v_i/t_i$ (dark green), and proceeds toward the central region.

Figure 7 demonstrates that different objective functions result in different selected samples. As seen in Table 1, the time-greedy method by design always maximizes the number of observed planets, although many of them share very similar characteristics (see the orange points in the top panel in Figure 7). In particular, planets which have large variance but require a lot of observational time are missed. On the other hand, the variance-greedy method selects all outliers, regardless of their observational cost, leading to a rather small (typically by a factor of 2) target sample (see the middle panel in Figure 7). In addition, many of the typical planets in the population are completely missed because the core of the distribution is not sampled. The leverage-greedy method (bottom panel in Figure 7) is trading some of the high-variance, high-cost planets for many more lower-variance, lower-cost planets, in order to strike the right balance between sample size and sample diversity. The resulting selection more evenly samples the original population, although the central (most typical) population is systematically underrepresented, reflecting the fact that planets near the mean contribute less to the leverage objective function. This behavior is intrinsic to leverage-based optimization and should be kept in mind when interpreting the resulting selection. Simulated annealing (lower right panel in Figure 8) tends to produce samples that are more representative of the underlying population, at the cost of a small reduction in leverage. The results in Table 1 also show that the leverage increases with the number of diversity axes approximately as $\sqrt{N_p}$, consistent with the definition of the objective function.

Figure 9 is an expansion of Figures 7 and 8 to show the properties of the selected planets (in color) across all three parameter axes. While the selection uses the planet temperature as the parameter of diversity (middle column), the left and the right columns show the distribution of the selected planets in terms of planetary radius and star temperature, respectively.

Figure 4 is a summary plot of the performance of all the different selection methods discussed in the paper for the one-parameter exercise; the corresponding results for the two- and three-parameter exercises are shown in Figures 5 and 6, respectively. Figures 4–6 illustrate the achieved leverage (color bar) of a given selection method on a scatter plot of sample variance versus sample size. Each point represents a possible sample selection and the green color contours outline samples of constant leverage. Each of the discussed sample-selection strategies is shown with a colored circle as indicated in the legend. It is clear that the time-greedy selection maximizes the number of selected planets given the mission time constraint, while the variance-greedy selection represents the other end of the spectrum, maximizing the variance of the selected sample. Both of these methods are outperformed in terms of leverage by the leverage-greedy selection. These figures demonstrate well that both the “regular-classes” and

**Table 1**
Comparison of Selection Approaches across 1D, 2D, and 3D Parameter Spaces

| Method | Planets Selected | Leverage Score | % of Budget Used |
|---|---|---|---|
| One parameter: Equilibrium temperature only | | | |
| Time greedy | **765** | 29.30 | 99.76% |
| Variance greedy | 326 | 32.99 | 99.94% |
| *Leverage greedy* | 530 | **34.08** | **100.00%** |
| *Regular classes* | 753 | 29.13 | 99.87% |
| *Quantile classes* | 761 | 28.38 | 99.84% |
| *K-means clustering* | 678 | 32.30 | 99.92% |
| *Simulated annealing* | 698 | 33.33 | 99.94% |
| Random selection | 333 | 17.05 | 100.00% |
| Two parameters: Planet radius + Equilibrium temperature | | | |
| Time greedy | **765** | 42.03 | 99.76% |
| Variance greedy | 344 | 45.81 | 99.80% |
| *Leverage greedy* | 551 | **47.63** | 99.99% |
| *Regular classes* | 678 | 41.23 | 99.99% |
| *Quantile classes* | 613 | 35.31 | 99.49% |
| *K-means clustering* | 678 | 44.13 | 99.92% |
| *Simulated annealing* | 695 | 45.80 | 99.93% |
| Random selection | 333 | 23.63 | **100.00%** |
| Three parameters: Planet radius + Equilibrium temperature + Stellar Temperature | | | |
| Time greedy | **765** | 52.06 | 99.76% |
| Variance greedy | 349 | 56.08 | 99.41% |
| *Leverage greedy* | 568 | **58.52** | 99.86% |
| *Regular classes* | 650 | 53.80 | 99.85% |
| *Quantile classes* | 502 | 34.31 | 99.39% |
| *K-means clustering* | 605 | 56.14 | 99.59% |
| *Simulated annealing* | 696 | 56.70 | 99.93% |
| Random selection | 333 | 28.99 | **100.00%** |

**Note.** The lines in italics denote the five different heuristic approaches to maximizing the leverage objective function $f_L$. We quote the number of planets selected, the leverage score, and the percentage of the time budget that is utilized, i.e., $\frac{1}{T}\sum_{k\in S} t_k$. For simplicity, here we report the leverage and not the leverage squared. Bold values are the biggest values for each case.

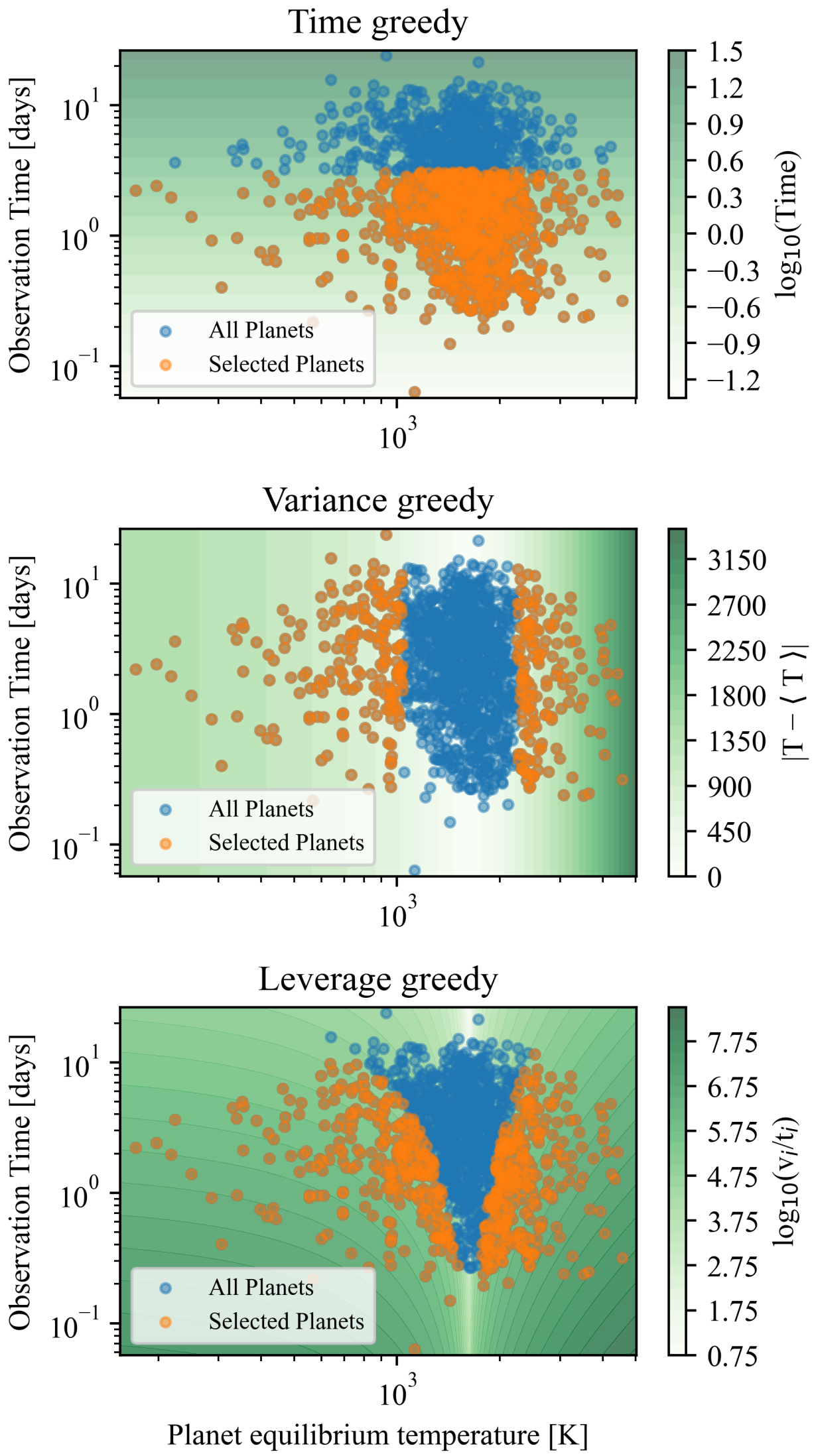

**Figure 7.** One-parameter exercise. Subpopulation of selected planets (in orange) using different selection objectives: sample size (Equation (2), top panel), sample variety (Equation (3), middle panel), and sample leverage (Equation (6), bottom panel). The color coding indicated on the color bar represents the contribution of an individual planet to the total time budget $T$ (top panel) or to the respective objective function (middle and lower panels).

"quantile-classes" selections perform very similar to the time-greedy method in terms of leverage and number of selected planets, in agreement with the conclusions in N. B. Cowan & B. Coull-Neveu (2025). It also makes clear that the simulated-annealing method gives very competitive results in terms of leverage, while it also allows for a large sample size. *K*-means clustering presents a viable alternative worth investigating, particularly if the dataset contains well-defined, distinct subgroups.

As a baseline for comparison, we also sample random subsets uniformly from all possible subsets of the original planet population which satisfy the time constraint (see Section 3.4). The random subsets (shown as black dots in Figures 4–6) tend to cluster in the bottom-left region of the plot, indicating the worst case: a low number of planets and a low variance. We can see that all optimization methods present a substantial improvement compared to the random subsets, which is reassuring. In other words, it pays off to have a global selection strategy, which motivates a community discussion about the science priorities and the mission constraints, having in mind the entire planet population rather than individually selected planets.

Figure 10 is a summary of the results from the two-parameter exercise that shows the distribution of the selected planets in the plane of the planet equilibrium temperature and planet radius. The panels correspond to the seven different selection methods discussed previously, now performed in two dimensions ($R_p$ and $T_p$). One can see that qualitatively the distribution of selected planets behaves similarly to the one-parameter case: e.g., the variance-greedy algorithm selects predominantly outliers away from the core of the original distribution.

Figure 11 presents cumulative curves, similar to Figure 3 in N. B. Cowan & B. Coull-Neveu (2025), of each selection method as a function of the observing time. The top row shows the cumulative number of targets as a function of the observing budget for the one-parameter, two-parameter, and three-parameter experiments. As expected, the time-greedy approach (blue line) is maximizing the number of observed planets and selects the most planets at the end of the observing cycle. The variance-greedy method (orange line) on the other hand, results in the smallest sample size as it does not prioritize the number of planets. One can observe notable "kinks" in the curves for regular classes (Red) and *K*-means clustering (magenta). The kinks occur because samples are not uniformly distributed across bins or clusters (see Figure 3). When a specific bin becomes empty, the selection algorithm must pivot to the remaining bins that contain more "accessible" (easier to observe) targets, causing a sudden shift in the curve's slope. The panels in the bottom row compare the cumulative leverage. As expected, the leverage-greedy and simulated-annealing strategies achieve a higher leverage value for a given observing time, indicating a better trade-off between diversity and efficiency. Class-based and clustering methods occupy an intermediate regime, offering structured coverage of the parameter space but lower total leverage.

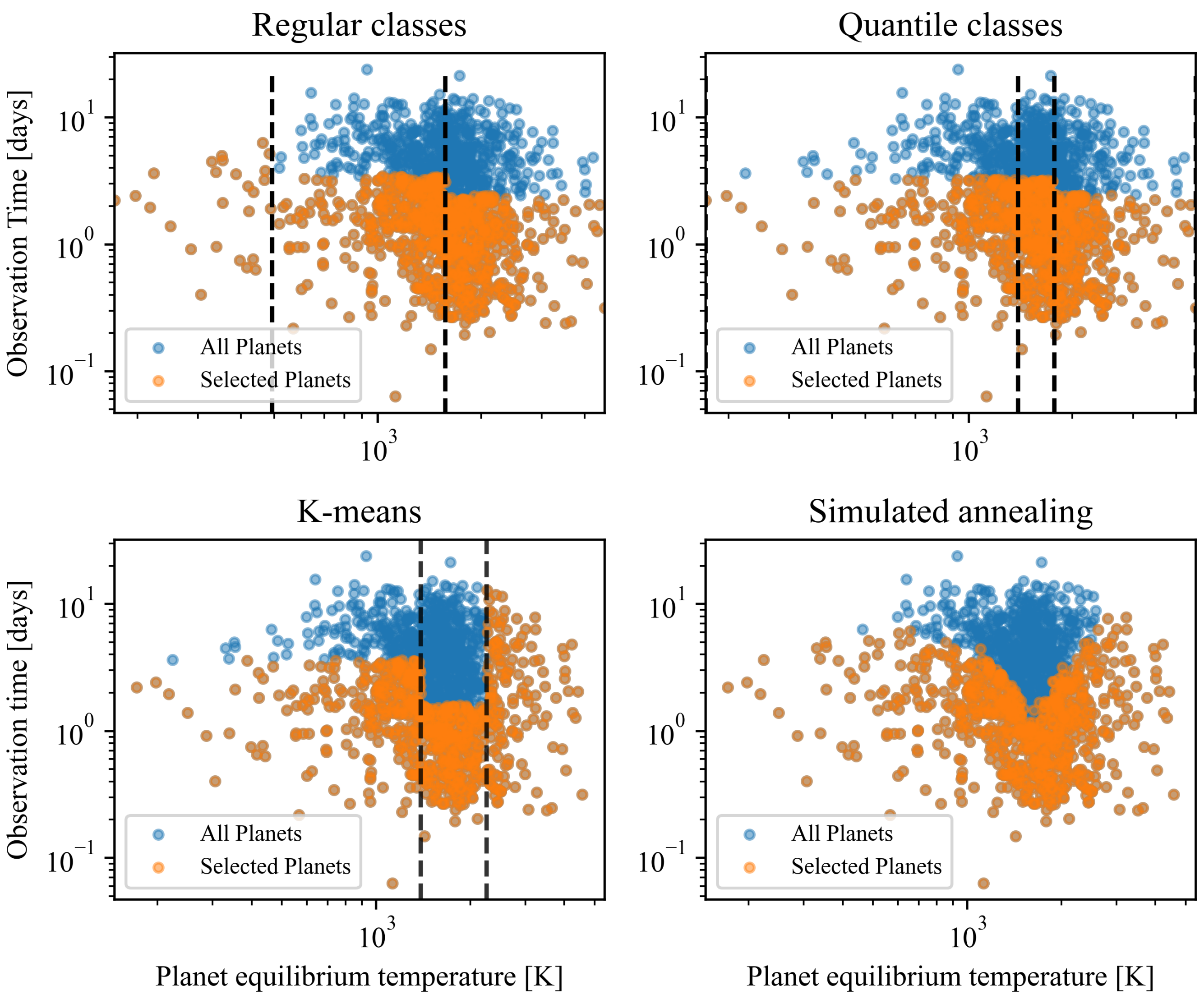

**Figure 8.** One-parameter exercise. Subpopulations of selected planets (in orange) for the leverage objective function (6), using alternative heuristics for maximizing the leverage objective function: regular classes (Section 3.3.1, top left panel), quantile classes (Section 3.3.2, top right panel), *K*-means clustering (Section 3.3.3, bottom-left panel), and simulated annealing (Section 3.3.5, bottom right panel). The selection using the leverage-greedy algorithm of Section 3.3.4 was already shown in the bottom panel of Figure 7.

The main objective in this paper is to compare the performance of different planet-selection strategies, subject to overall budget constraints. In this work, we did not take into account any observational uncertainties on the planet/star parameters, which in practice may affect the scientific return in different ways. First, uncertainties in the system parameters used for binning can lead to incorrect class designation—objects near bin boundaries could move between categories. Second, selecting planets with large uncertainties is undesirable since it reduces confidence in the resulting scientific interpretation. Taking uncertainties into account is particularly important for planet outliers, as they can disproportionally impact the selected sample. To add to the complexity of the problem, the uncertainties among the different parameters for a single planet could differ wildly. We recognize that this is an important problem and reserve that for future work. A subsequent analysis would incorporate the observational uncertainties into the objective functions considered here.

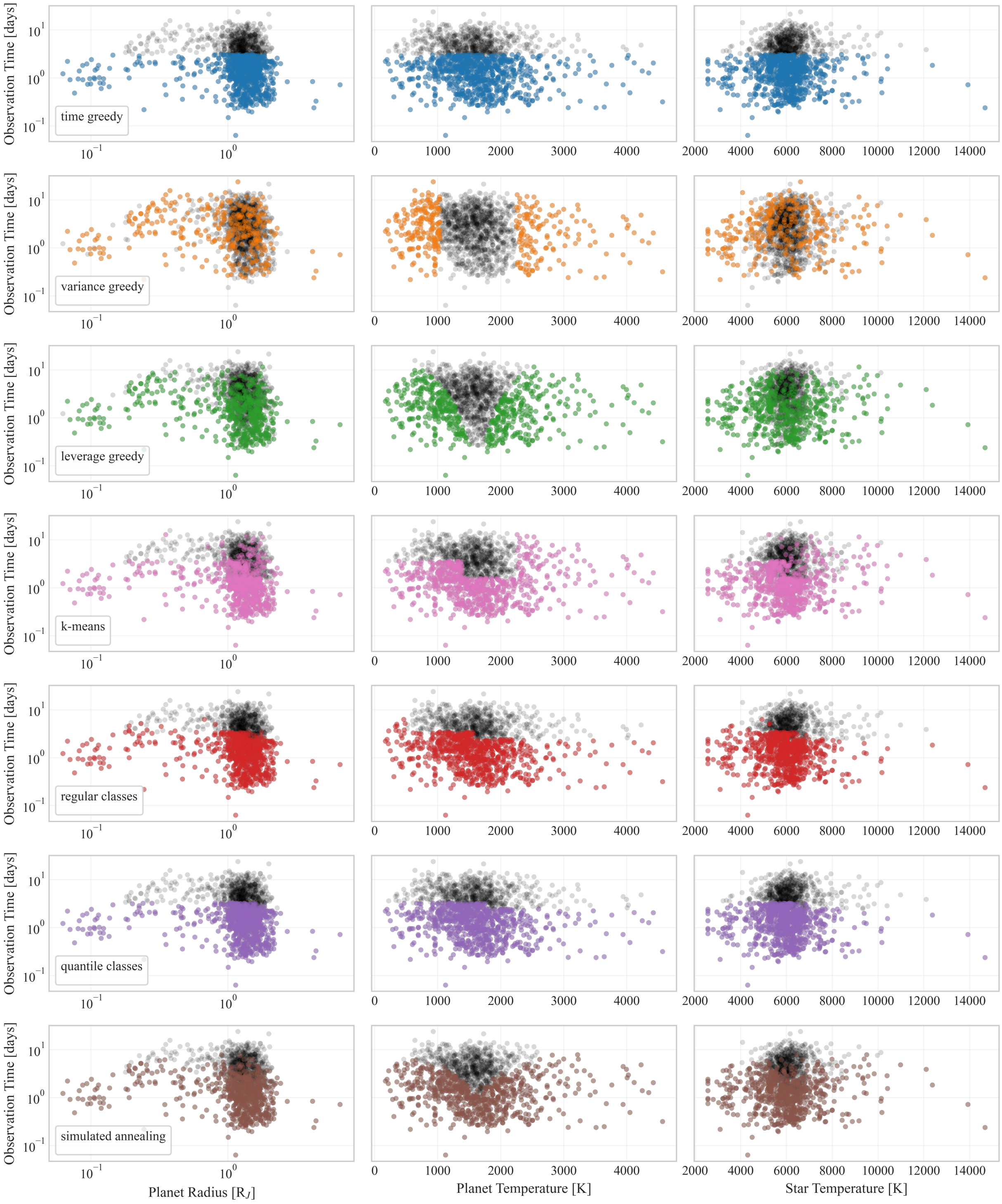

**Figure 9.** One-parameter exercise. Subpopulations of selected planets (in color) for the first seven methods listed in Table 1, as labeled on the panels in the left column. The selection uses the planet temperature as the parameter of diversity (middle column), while the left and the right columns show the distribution of the selected planets in terms of planetary radius and star temperature, respectively.

## 5. Summary and Conclusions

In this work we compared several approaches for selecting exoplanet targets under an observation time constraint. We introduced three different selection criteria (sample size, sample variance, and sample leverage) and quantified them in terms of suitable objective functions, allowing for the application of modern optimization algorithms. For maximizing the leverage objective function we considered five different heuristics (leverage greedy, simulated annealing, *K*-means clustering, regular classes, and quantile classes). The outcomes from the different selection methods were illustrated with three exercises (each with one, two, or three axes of diversity, respectively) and the results were collected in Table 1 and Figures 4–11.

Each selection criterion highlights a different compromise between the number of planets selected, their diversity, and the total observation time required. The time-greedy selection maximizes the number of planets but not the diversity. The variance-greedy selection maximizes diversity, but that comes at a cost: it overemphasizes rare cases, undersamples the most typical planets, and, in general, sacrifices the number of planets which can be observed. The leverage-greedy selection appears to be a good compromise for balancing sample diversity and sample size. In this sense, leverage-based selection provides a strong strategy for surveys aiming to characterize linear population-level trends under limited time, such as Ariel.

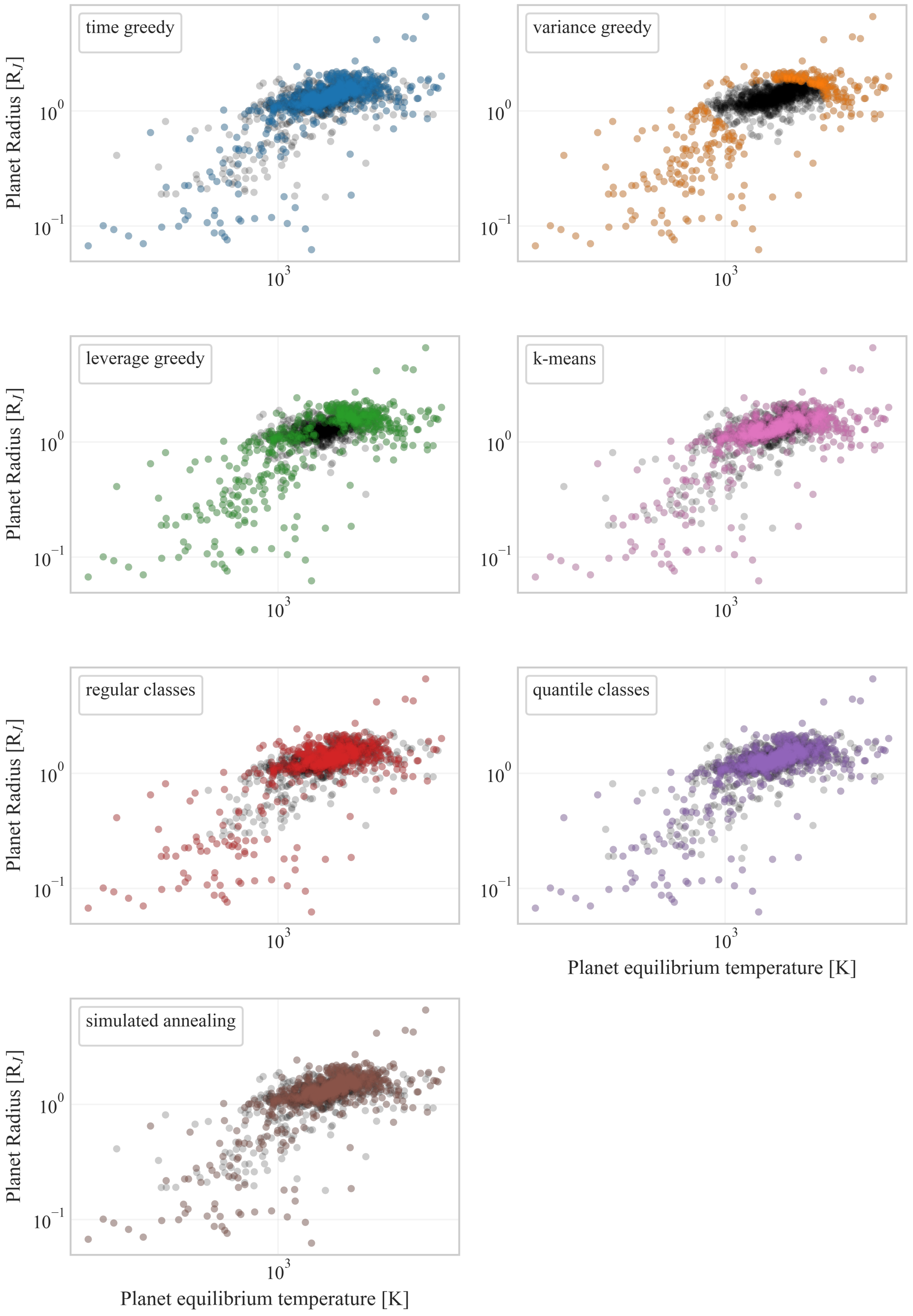

**Figure 10.** Scatter plots in the plane of the equilibrium temperature and the planet radius of the total population of planets (in black) and the planets selected by the different methods discussed in the text (in color).

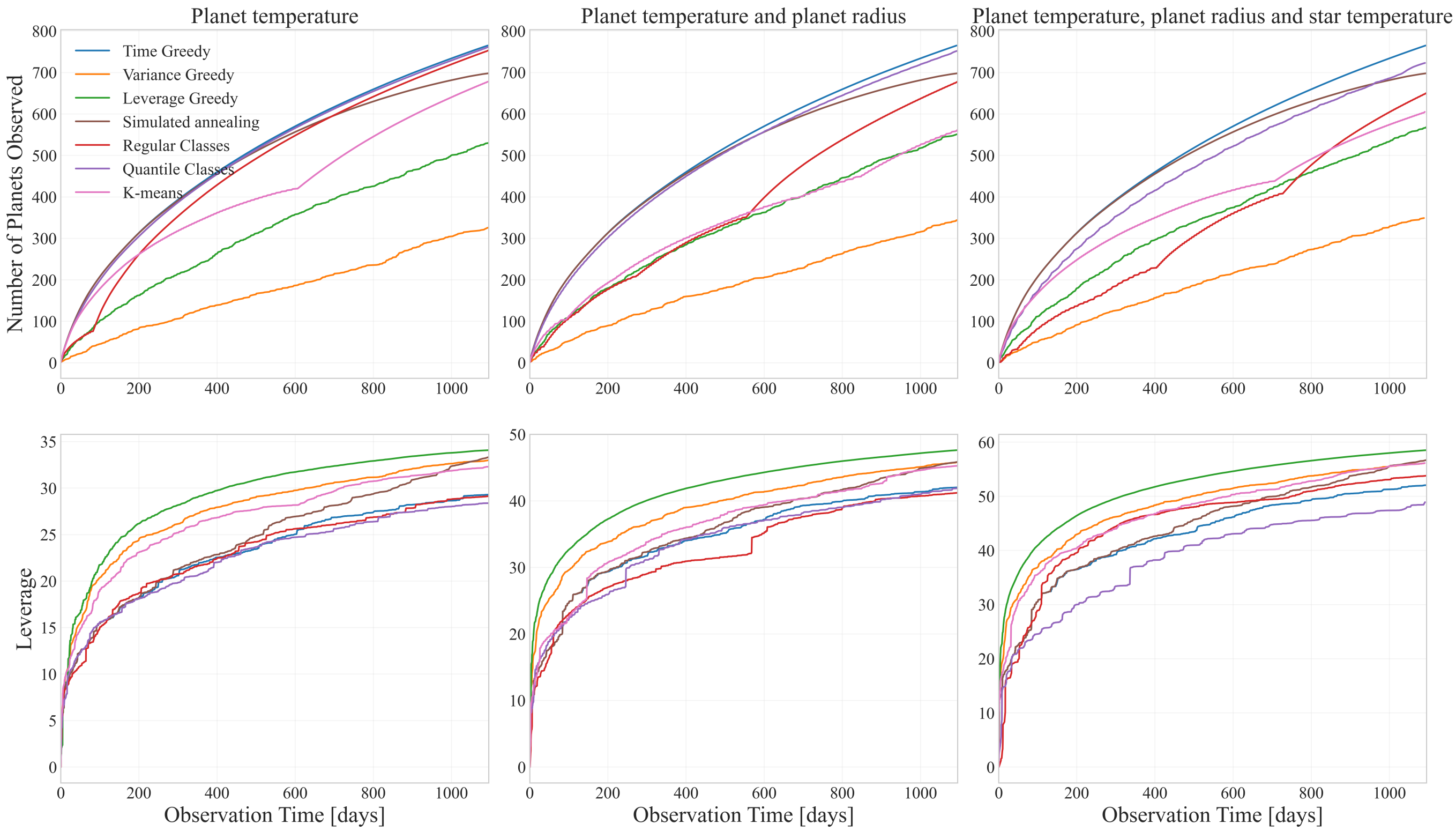

**Figure 11.** Comparison of cumulative curves for each of the seven selection strategies discussed in the paper. The top row shows the number of planets selected as a function of observation time in days. The bottom row similarly shows the sample leverage as a function of observation time. The left (middle, right) columns represent the results for the one- (two-, three-) parameter exercise, respectively.

Table 1 and Figures 4–6 demonstrate that among the five heuristics for optimizing the leverage, the leverage-greedy selection achieves the highest leverage value. As discussed in Section 4, simulated annealing is always second best and provides an excellent alternative, with a significantly increased sample size for a very modest reduction in leverage. In contrast to the leverage-greedy selection, the selection with simulated annealing is much more representative of the original population and covers both rare and typical planets (see the lower right panel in Figure 8). Binning or clustering methods ensure structured coverage of the entire parameter space.

We note that the use of standardized variables (17) allows us to easily generalize the discussion to several axes of diversity, as shown in Figures 5, 6, 10, and 11. Furthermore, it helps us understand the scaling of the sample leverage with a different number of axes of diversity (i.e., parameters of interest $N_p$). Note that the number of terms in the sum (16) representing the leverage *squared* scales with $N_p$, therefore the leverage itself scales with $\sqrt{N_p}$. This can be easily verified by the results in Table 1: the maximal leverage obtained in the one-parameter exercise was ~34, so for the two-parameter exercise we should expect $\sim 34 \times \sqrt{2} \sim 48$, and that is precisely what is seen in the middle section of Table 1. Similarly, for the three-parameter exercise we would expect the maximal leverage to be $\sim 34 \times \sqrt{3} \sim 58.9$, again in agreement with the lower section of Table 1.

In all examples considered in this paper, we always used a single selection criterion (objective function) throughout the selection process. One possible extension of this work would be to combine multiple strategies. For example, we could first use leverage to secure diversity in a small subset of targets, while using another criterion (or a list of specific targets of interest) for the remainder of the selection. That would ensure that unusual planets are represented, while also having enough typical planets or planets of interest selected.

From a practical perspective, our results suggest that methods which explicitly optimize leverage are best suited for maximizing the constraints on population-level trends. At the same time, alternative approaches such as simulated annealing may offer a useful compromise by improving the representativeness of the selected sample with only a modest reduction in leverage. The choice between these strategies should therefore be guided by whether the primary goal is statistical efficiency or broader coverage of the underlying population.

As discussed in N. B. Cowan & B. Coull-Neveu (2025), the leverage is directly related to the uncertainty on the slope of linear trends. Many interesting relationships in planetary science and astrophysics are linear (or power laws which can be equivalently transformed into linear). Our analysis can be easily extended to other, nonlinear relationships by simply designing the appropriate heuristics as we demonstrated in this paper. In future work, we plan to apply our approach to the mass–metallicity relation for exoplanets, as it is one of the most fundamental population-level trends noted in the literature (D. P. Thorngren et al. 2016; L. Welbanks et al. 2019; Q. Sun et al. 2024; M. R. Swain et al. 2024). This will provide a physically motivated test case linking optimization of target selection to potential scientific results of the Ariel mission. We also plan to incorporate ephemerides and scheduling constraints, as the periodic observability of exoplanets critically affects the selection of targets. Taking into account these factors will bring the optimization closer to real mission planning.

## Acknowledgments

The work of E.P., A.R., and K.T.M. is supported in part by the Shelby Endowment for Distinguished Faculty at the University of Alabama. N.B.C. acknowledges support from an NSERC Discovery Grant and a Tier 2 Canada Research Chair and thanks the Trottier Space Institute and l'Institut de recherche sur les exoplanètes for their financial support and dynamic intellectual environment. The authors thank Ben Coull-Neveu for sharing his codes from the previous related study in N. B. Cowan & B. Coull-Neveu (2025). The authors thank Élise Leclerc for carefully reading the manuscript and identifying an error.

## ORCID iDs

Emilie Panek https://orcid.org/0009-0000-8050-5348
Alexander Roman https://orcid.org/0000-0003-2719-221X
Katia Matcheva https://orcid.org/0000-0003-3074-998X
Konstantin T. Matchev https://orcid.org/0000-0003-4182-9096
Nicolas B. Cowan https://orcid.org/0000-0001-6129-5699